\documentstyle[A4,11pt,epsfig,twoside]{article}
\def\st{\mbox{$\rm \tilde{t}_1$}}
\def\s2{\mbox{$\rm \tilde{t}_2$}}
\def\stl{\mbox{$\rm \tilde{t}_L$}}
\def\sr{\mbox{$\rm \tilde{t}_R$}}
\def\mt{\mbox{$m_{\rm \tilde{t}_1}$}}
\def\m2{\mbox{$m_{\rm \tilde{t}_2}$}}
\def\cost{\mbox{$\cos\theta_{\rm \tilde t}$}}
\def\sint{\mbox{$\sin\theta_{\rm \tilde t}$}}
\def\ee{\mbox{$\rm e^+e^-$}}

\def\ra{\rightarrow}
\def\qq{\mbox{$\rm q\bar q$}}
\def\mco{\multicolumn}

\begin{document}
\begin{titlepage}
\def\thefootnote{\fnsymbol{footnote}}       

\begin{center}
\mbox{ } 

\end{center}
\vskip -3.0cm
\begin{flushright}
\Large
\vspace*{-5cm}
\mbox{\hspace{11.85cm} hep-ph/0211140} \\
\mbox{\hspace{12.0cm} November 2002}
\end{flushright}
\begin{center}
\vskip 1.0cm
{\boldmath \Huge\bf
{Precision Measurements in the 

Scalar Top Sector of the MSSM 
\smallskip
at a Linear $\rm e^+e^-$ Collider}
}
\vskip 1cm
{\LARGE\bf A.~Finch$^1$, H.~Nowak$^2$ and A.~Sopczak$^1$\\
\smallskip
\smallskip
\smallskip
\Large
$^1$ Lancaster University, UK, \\
$^2$ DESY Zeuthen, Germany}

\vskip 1.5cm
\centerline{\Large \bf Abstract}
\end{center}

\vskip 3.cm
\hspace*{-2cm}
\begin{picture}(0.001,0.001)(0,0)
\put(,0){
\begin{minipage}{17cm}
\Large
\renewcommand{\baselinestretch} {1.2}
The scalar top discovery potential has been studied 
for the TESLA project
with a full-statistics background simulation at $\sqrt{s}=500$~GeV
and ${\cal L}=500$~fb$^{-1}$.
The beam polarization is very important to measure the scalar top 
mixing angle and to determine its mass.
The latest estimation of the beam polarization parameters is applied.
This study includes $\rm e^+$ polarization, which improves the
sensitivity.
For a 180 GeV scalar top at minimum production cross section, 
we obtain 
$\Delta m=0.8$~GeV and $\Delta\cost=0.008$ in the
neutralino channel,
and $\Delta m=0.5$~GeV and $\Delta\cost=0.004$
in the chargino channel.
The MSSM parameters for the Snowmass point 5 at $\cost=0.513$ predict a
scalar top of 210 GeV and a neutralino of about 120 GeV.
For this parameter choice $\Delta m=0.95$~GeV and $\Delta\cost=0.0125$
is obtained, 
and a preliminary c-tagging study is presented in the neutralino channel.
\renewcommand{\baselinestretch} {1.}

\normalsize
\vspace{2cm}
\begin{center}
{\sl \large
Presented at the LCWS02, Korea, Aug. 2002
\vspace{-3cm}
}
\end{center}
\end{minipage}
}
\end{picture}
\vfill

\end{titlepage}


\newpage
\thispagestyle{empty}
\mbox{ }
\newpage
\setcounter{page}{1}

\title{\mbox{PRECISION MEASUREMENTS IN THE SCALAR TOP}
       \mbox{SECTOR OF THE MSSM AT A LINEAR $\rm e^+e^-$ COLLIDER}}

\author{A.~FINCH$^1$, H.~NOWAK$^2$ and A.~SOPCZAK$^1$\footnote{speaker}\\
\smallskip
$^1$Lancaster University, UK,~~~ $^2$DESY Zeuthen, Germany}

\date{}
\maketitle

\vspace*{-1.5cm}
\begin{abstract}
\vspace*{-0.5cm}
The scalar top discovery potential has been studied 
for the TESLA project
with a full-statistics background simulation at $\sqrt{s}=500$~GeV
and ${\cal L}=500$~fb$^{-1}$.
The beam polarization is very important to measure the scalar top 
mixing angle and to determine its mass.
The latest estimation of the beam polarization parameters is applied.
This study includes $\rm e^+$ polarization, which improves the
sensitivity.
For a 180 GeV scalar top at minimum production cross section, 
we obtain 
$\Delta m=0.8$~GeV and $\Delta\cost=0.008$ in the
neutralino channel,
and $\Delta m=0.5$~GeV and $\Delta\cost=0.004$
in the chargino channel.
The MSSM parameters for the Snowmass point 5 at $\cost=0.513$ predict a
scalar top of 210 GeV and a neutralino of about 120 GeV.
For this parameter choice $\Delta m=0.95$~GeV and $\Delta\cost=0.0125$
is obtained, 
and a preliminary c-tagging study is presented in the neutralino channel.
\end{abstract}

\vspace*{-0.9cm}
\section*{Introduction}
\vspace*{-0.6cm}

The study of the scalar top quarks is of particular interest, since the
lighter stop mass eigenstate is likely to be the lightest scalar quark in a 
supersymmetric theory. The mass eigenstates are \mt\ and \m2\ with 
$\mt < \m2$, where $\st=\cost \stl + \sint \sr$ and 
$\s2=-\sint \stl + \cost \sr$ with the mixing angle \cost.
We study the experimental possibilities to
determine \mt\ and \cost\ at a high-luminosity \ee\ linear collider 
such as the TESLA project~\cite{tesla} with polarized
$\rm e^+$ and $\rm e^-$ beams.

The simulated production process is 
$\ee\rightarrow\st\bar{\st}$ with two decay modes
$\st\rightarrow\tilde{\chi}^0$c and 
$\st\rightarrow\tilde{\chi}^+$b.
A 100\% branching fraction  in each decay mode is simulated.
The first scalar top decay into a c-quark and the lightest neutralino 
results in a signature of two jets and large missing energy.
The second investigated stop decay mode leads also to large missing energy
and further jets from the chargino decay.
The neutralino channel is dominant unless the decay into a chargino is
kinematically allowed.
Details of the event simulation with SGV~\cite{sgv} tuned for a 
TESLA detector~\cite{tesla} are given in Ref.~\cite{epj}.
The signals and a total of 16 million Standard Model 
background events are simulated (Table~\ref{tab:pre})
for ${\cal L}=500$~fb$^{-1}$.

\begin{table}[hp]
\vspace*{-0.4cm}
\begin{center}
\begin{tabular}{c|c|c|c|c|c|c|c|c}
Channel      & $\tilde{\chi}^0$c$\tilde{\chi}^0\mathrm{\bar{c}}$
             & $\tilde{\chi}^+$b$\tilde{\chi}^-\mathrm{\bar{b}}$
             &eW$\nu$  & WW & $\rm q\bar{q}$   
             & t$\rm \bar{t}$&ZZ   & eeZ   \\ \hline
(in 1000)    & 50   & 50 & 2500 & 3500  &6250  & 350 & 300 & 3000 \\ 
\end{tabular}
\vspace*{-0.6cm}
\end{center}
\caption{\label{tab:pre} 
Number of simulated signal and background events.}
\end{table}

\vspace*{-0.8cm}
\section*{Neutralino Channel}
\vspace*{-0.6cm}

The reaction $\rm \ee\rightarrow\st\bar{\st}\rightarrow 
\tilde{\chi}^0c\tilde{\chi}^0\bar{c}
$ has been studied for a 180 GeV scalar top and a 100 GeV neutralino.
After a preselection,
278377 background events remain~\cite{sitgesproc,epj}.
In order to separate the signal from the background, 
the following selection variables are defined: 
visible energy,
number of jets, 
thrust value and direction, 
number of clusters, 
transverse and parallel momentum imbalance, 
acoplanarity and invariant mass of two jets~\cite{sitgesproc,epj}.
An Iterative Discriminant Analysis (IDA)~\cite{ida} optimized 
the selection. For unpolarized beams and 
12\% efficiency, 400 background events are expected.

The polarization of the $\rm e^+$ and $\rm e^-$ beams at a future 
linear collider offers the opportunity to enhance or suppress 
the left- or right-handed couplings 
of the scalar top signal and to determine mass and 
mixing angle independently. The production cross section of each background 
process depends differently on the polarization. It is therefore important 
for a high-statistics analysis to study the expected background channels 
individually.
The expected cross sections~\cite{xsec,generator} are given in 
Table~\ref{tab:bgxsec} for different beam polarization states.
The IDA was repeated for $-0.9$ and $0.9$ 
polarization~\cite{epj}\,\footnote{For a polarization of 
$-0.9$, 95\% of the $\rm e^-$ are left-polarized.
In the previous analyses~\cite{moriokaproc,desy123d,desy123e,zphys}
it was assumed that only 90\% of the $\rm e^-$ were polarized.}.
We recalculated all background rates for the new machine polarization 
of $-0.8 / 0.6$ (left-polarization) and $0.8 / -0.6$ (right-polarization) 
as a function of the signal efficiency~\cite{ashn01}.
For 12\% detection efficiency, 1194 background events are expected
leading to $\sigma_{\rm left} = 81.8 \pm 1.3 $ fb, 
and 208 background events giving $\sigma_{\rm right} = 76.4 \pm 1.2 $ fb, 
where $\Delta\sigma/ \sigma = 
\sqrt{N_{\rm signal} + N_{\rm background}} / N_{\rm signal}$.

\begin{table}[tp]
\vspace*{-0.5cm}
\begin{center}
\begin{tabular}{c||c|c|c|c|c|c|c} 
 Pol.  &Pol. & $\st\bar{\st}$ &$\rm W e \nu$ &  WW & $\rm q\bar q$ &$\rm t\bar t$ &  ZZ  \\
 of $\rm e^-$ & of $\rm e^+$ 
&{\small \hspace*{-1mm}CALVIN}\hspace*{-1mm}     
&{\small \hspace*{-1mm}GRACE}\hspace*{-1mm}     
&{\small \hspace*{-1mm}WOPPER}\hspace*{-1mm}   
&{\small \hspace*{-1mm}HERWIG}\hspace*{-1mm} 
&{\small \hspace*{-1mm}HERWIG}\hspace*{-1mm}
&{\small \hspace*{-1mm}COMPHEP}\hspace*{-1mm} \\ \hline
  $-0.8$& 0.6    & 0.0818 & 10.7  & 22.6 & 21.5  & 1.11 & 0.909 \\
  $-0.9$& 0.0    & 0.0552 & 6.86  & 14.9   & 14.4  & 0.771  & 1.17  \\
  0     & 0.0    & 0.0535 & 5.59  &  7.86  & 12.1  & 0.574  & 0.864 \\
  0.9   & 0.0    & 0.0517 & 4.61  &  0.906 &  9.66 & 0.376  & 0.554 \\ 
  0.8   & $-0.6$ & 0.0764 & 1.78 & 0.786 & 13.0 & 0.542 & 0.464  \\ 
\end{tabular}
\end{center}
\vspace*{-0.5cm}
\caption{\label{tab:bgxsec} 
Signal and background cross\,sections (pb) from different
event generators for $\rm e^-$ and $\rm e^+$ polarization states
($\mt=180$~GeV and $\cost=0.57$).
The Zee cross section is 0.6~pb.}
\vspace*{-0.5cm}
\end{table}

\vspace*{-0.8cm}
\section*{Chargino Channel}
\vspace*{-0.6cm}

The reaction $\rm \ee\rightarrow\st\bar{\st}\rightarrow 
\tilde{\chi}^+b\tilde{\chi}^-\bar{b} \rightarrow
\tilde{\chi}^0W^+b\tilde{\chi}^0W^-\bar{b}$ 
has been studied with focus on the hadronic W decays
for a 180 GeV scalar top, a 150 GeV chargino and a 60 GeV neutralino, 
where the chargino decays 100\% into a W boson and a neutralino.
A preselection  similar to that for the study of the neutralino channel
is applied and 209051 background events remain~\cite{epj}.
In order to separate the signal from the background, 
the following selection variables are defined: 
visible energy,
number of jets,
thrust value,
number of clusters,
transverse and parallel momentum imbalance,
and the isolation angle of identified leptons.
For an unpolarized beam, the final IDA 
output variable and the resulting number of background events as a 
function of the signal efficiency were calculated~\cite{ashn01}.
For 12\% efficiency, only 20 background events are expected.
Allowing a background rate of 400 events, as in the neutralino channel,
the efficiency is 44\%, from which we derive the relative error
on the cross section to be 0.75\%.
In this case the number of expected signal events is much larger
than the expected background, thus no separate tuning of the IDA for
left- and right-polarization is required.
The background is neglected for the determination of mass and mixing 
angle. Note that the dependence of the background rates on the polarization
has to be taken into account for stop masses closer to the kinematic 
threshold.

\vspace*{-0.8cm}
\section*{Snowmass Point 5} 
\vspace*{-0.6cm}

At the Snowmass 2001 ''Summer Study on the
Future of Particle Physics" a consensus was reached about a number of
defined mSugra points. One of them is dedicated to a light scalar top
\cite{snowmasspoints}. For this point the parameters $m_0=150$~GeV,
$m_{1/2}=300$~GeV, $A_0=-1000$, $\tan\beta =5$~ and~ $\mu > 0 $
generate a light scalar top at 210 GeV with a mixing angle
of $\cost=0.513$ and a neutralino of 121.2 GeV mass (SUSYGEN~\cite{susygen}). 
The scalar top decays
at 100\% into $\rm \tilde{\chi}^0 c$ because the chargino mass is about 
10~GeV higher than the scalar top mass. 
A three-particle decay of the scalar top into 
b--lepton--sneutrino is also excluded because the sneutrino mass is too high.
The cross section for the production of a scalar top at 210 GeV and
$\cost=0.513$ is roughly half of that obtained for a 180 GeV scalar top
at minimum cross section. As the mass differences between scalar top and
neutralino for both scenarios are roughly the same (80 and~ 90~ GeV,
respectively)  the same Standard Model backgrounds have been taken into
account. Thus, only the number of signal events decreases by about a factor 
of two.

In this detailed scenario light MSSM particles are
the 220~GeV scalar top 1, 
the 121~GeV neutralino 1,
the 223~GeV neutralino 2,
the 223~GeV chargino 1,
and the 114~GeV Higgs bosons.
Therefore, possible SUSY background reactions are:
\begin{center}
\vspace*{-0.4cm}
\begin{tabular}{lcl|lcl|lcl}
\mco{3}{c|}  {$\rm\ee\ra\tilde{\chi}^0_2\tilde{\chi}^0_1 \ra Z \tilde{\chi}^0_1 \tilde{\chi}^0_1$} 
&\mco{3}{|c|}{$\rm\ee\ra\tilde{\chi}^0_2\tilde{\chi}^0_2 \ra Z \tilde{\chi}^0_1 Z \tilde{\chi}^0_1$} 
&\mco{3}{|c} {$\rm\ee\ra\tilde{\chi}^+\tilde{\chi}^-\ra W \tilde{\chi}^0_1 W \tilde{\chi}^0_1$} \\\hline
(1a) & $\ra$& $\rm \qq \tilde{\chi}^0_1 \tilde{\chi}^0_1$&(2a) &$\ra$&$\rm \qq \tilde{\chi}^0_1 \qq \tilde{\chi}^0_1$&(3a) &$\ra$&$\rm \qq \tilde{\chi}^0_1 \qq \tilde{\chi}^0_1$  \\
(1b) & $\ra$& $\rm \ell^+\ell^- \tilde{\chi}^0_1 \tilde{\chi}^0_1$  &(2b) &$\ra$&$\rm \qq \tilde{\chi}^0_1 \ell^+\ell^- \tilde{\chi}^0_1$&(3b) &$\ra$&$\rm \qq \tilde{\chi}^0_1 \ell \nu \tilde{\chi}^0_1$ \\
\end{tabular} 
\end{center}
Only (1a) has a signature similar to the signal, 
however, the expected production rate is very small for this channel 
as $\rm \sigma(\ee\ra\tilde{\chi}^0_1\tilde{\chi}^0_2) = 0.037$~pb and
the decay branching ratios are $\rm BR(\tilde{\chi}^0_2\ra Z~\tilde{\chi}^0_1) = 0.012$ and
$\rm BR(Z\ra \qq) = 0.70$.
Thus, 
$\rm \sigma(\ee\ra\tilde{\chi}^0_1\tilde{\chi}^0_2\ra 
\qq\tilde{\chi}^0_1\tilde{\chi}^0_1) = 0.31$~fb 
and this background rate is expected to be much smaller than the 
expected signal rate.
Therefore, no significant background from Supersymmetric processes is
expected.

\vspace*{-0.8cm}
\section*{C-tagging with SIMDET-4} 
\vspace*{-0.6cm}

The 
preselection~\cite{epj} 
has been applied again on generated signal and background events 
after the events have passed a new detector simulation 
(SIMDET-4~\cite{simdet}) in order to study the c-tagging performance.
The vertex simulation is based on the TESLA CCD vertex 
detector from the LCFI Collaboration~\cite{lcfi}.
For each of the two reconstructed jets a value between 
0 and 1 is assigned according to the likelihood of it being
a charm jet. The product of the these numbers gives a variable 
that tags whether the event contains charm quarks or not.
After preselection cuts~\cite{epj} 4841/10k signal, 
9450/100k $\rm We\nu$ and 589/100k \qq\ remain. 
The production cross sections for no polarization were 
applied.
The preliminary efficiency versus purity is shown in 
Fig.~\ref{fig:point5} for the 
$\rm \tilde{\chi}^0c\tilde{\chi}^0\bar c$ channel
and the \qq\ and $\rm We\nu$ background reactions.
Purity is defined as the 
ratio of number of simulated signal events after the selection
to all selected events.
The c-tagging will further improve the final signal to 
background ratio and help to distinguish a scalar top 
signal from other possible Supersymmetric reactions 
involving jets and missing energy.

\vspace*{-0.8cm}
\section*{Results}
\vspace*{-0.6cm}

\begin{figure}[hb]
\vspace*{-1.0cm}
\begin{center}
\mbox{\epsfig{file=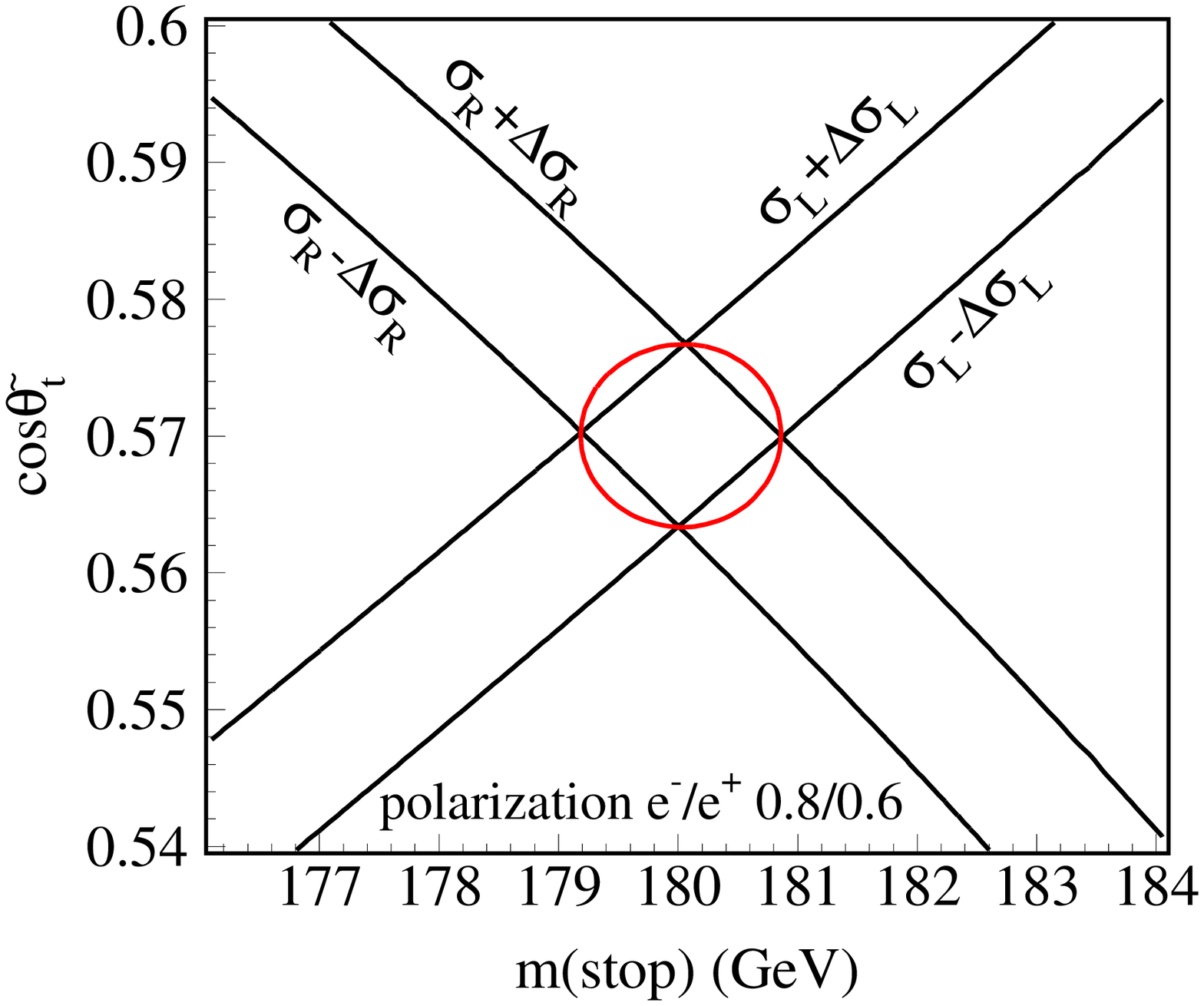,width=0.49\textwidth}}
\mbox{\epsfig{file=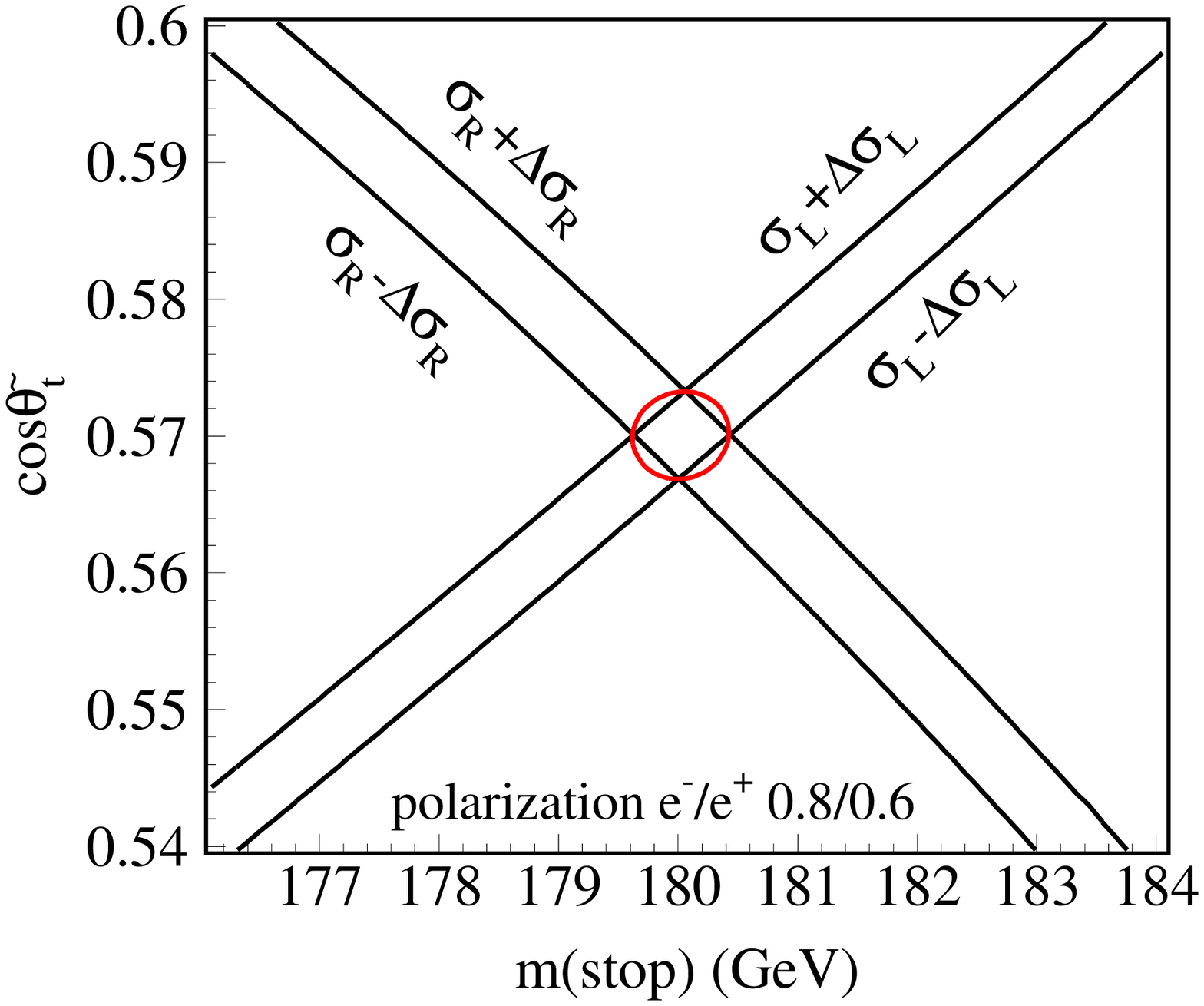,width=0.49\textwidth}}
\end{center}
\vspace*{-1.0cm}
\caption{\label{fig:ellipse} 
Error bands and the corresponding error ellipse 
as a function of \mt\ and \cost\ for $\sqrt s =500$~GeV,
${\cal L}=500$~fb$^{-1}$, $\mt=180$~GeV and $\cost=0.57$.
Left: neutralino channel. Right: chargino channel.}
\end{figure}

For the neutralino and chargino channels of scalar top quarks,
we have determined the expected Standard Model background rates 
as a function of the signal efficiency.
The total simulated background of about 16 million events is largely reduced,
which allows a precision measurement of the scalar 
top production cross section with a relative error of better than 2\%
in the neutralino channel and about 1\% in the chargino channel.
Based on experiences gained at LEP, we expect that detection efficiencies 
for other mass combinations are similar as long as the mass difference 
between the scalar top and the neutralino is larger than about 20 GeV.
Figure~\ref{fig:ellipse} shows the corresponding error bands and the error
ellipse in the \mt\ -- \cost\ plane for both decay channels
for $0.8/0.6$ left- and right-polarization of the $\rm e^-/e^+$ beams
and a luminosity of 500 fb$^{-1}$ each.
The statistical errors are a factor 7 better in the neutralino channel and
about a factor 14 better in the chargino channel than reported
previously~\cite{moriokaproc,desy123d,desy123e,zphys}, and improve further 
when in addition e$^+$ polarization is included. 
Detailed results are given in Table~\ref{tab:sum}.
The highest statistical precision is obtained in the chargino channel 
with an error $\Delta\mt=0.4$~GeV for $\mt=180$~GeV and
$\Delta\cost=0.003$ $\cost=0.570$.
\begin{table}[h!]
\vspace*{-0.5cm}
\begin{center}
\begin{tabular}{c|c|c||cc|cc} 
$\cal L$ (fb$^{-1}$) & e$^-$ Pol. & e$^+$ Pol. 
& (a)  $\Delta\mt$ & $\Delta\cost$ & (b) $\Delta\mt$ & $\Delta\cost$ \\\hline
10  & 0.8  &  0.0  &  7.0 & 0.06  & 7.0  & 0.06  \\
500 & 0.9  &  0.0  &  1.0 & 0.009 & 0.5  & 0.004 \\
500 & 0.8  &  0.6  &  0.8 & 0.008 & 0.4  & 0.003 \\
\end{tabular}
\end{center}
\vspace*{-0.5cm}
\caption{\label{tab:sum}
Expected errors on the scalar top mass and mixing angle from simulations
with different luminosity and beam polarization
in the neutralino (a) and chargino (b) channels.
The 10 fb$^{-1}$ 
analysis~\protect\cite{moriokaproc,desy123d,desy123e,zphys}
used a sequential event selection; while the 500~fb$^{-1}$ results in the
neutralino~\protect\cite{sitgesproc} and 
chargino~\protect\cite{epj} channels
are based on an IDA. The new result includes e$^+$ polarization.}
\vspace*{-0.5cm}
\end{table}

\begin{figure}[hb]
\vspace*{-0.2cm}
\begin{center}
\mbox{\epsfig{file=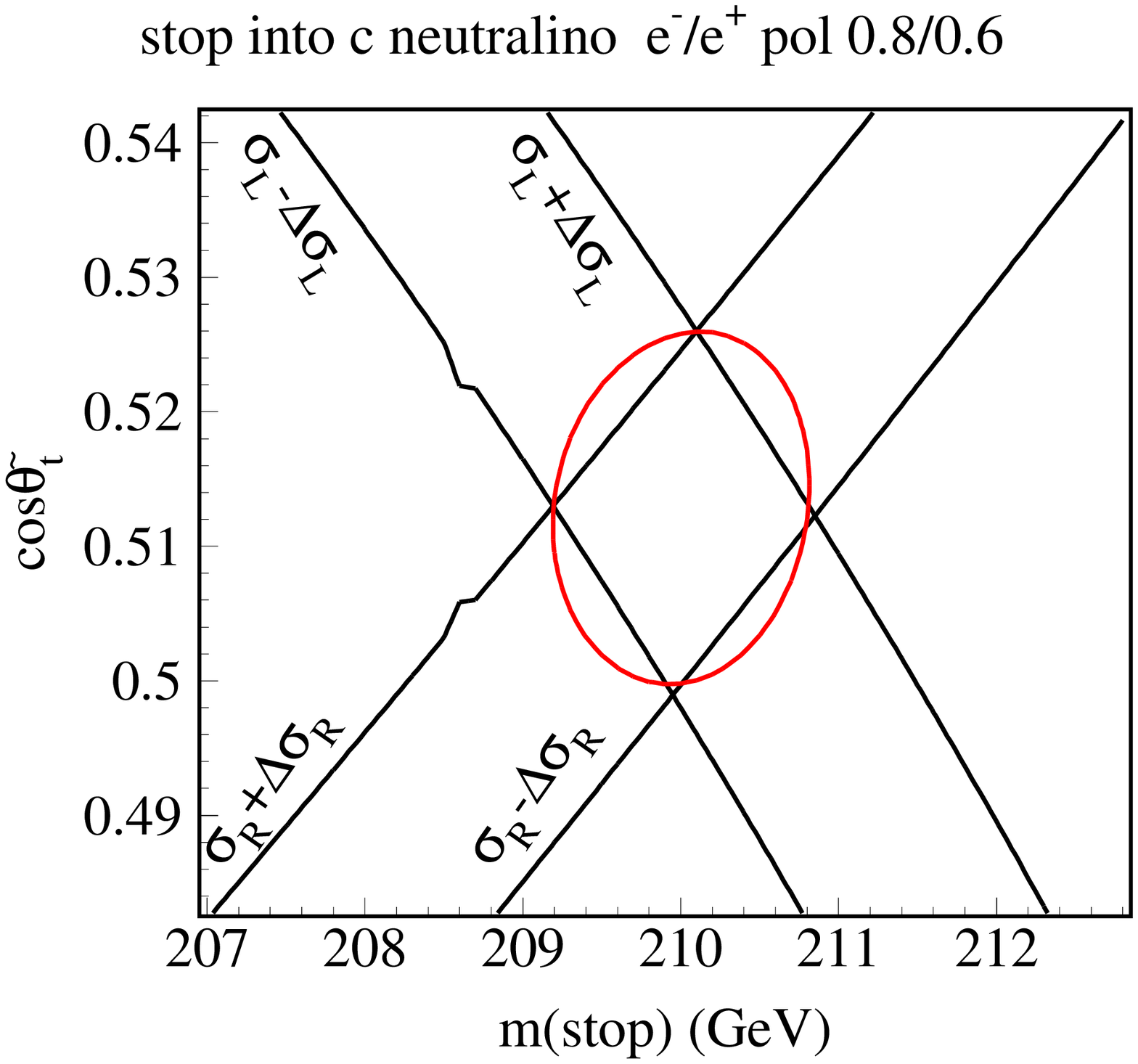,width=0.49\textwidth}}
\mbox{\epsfig{file=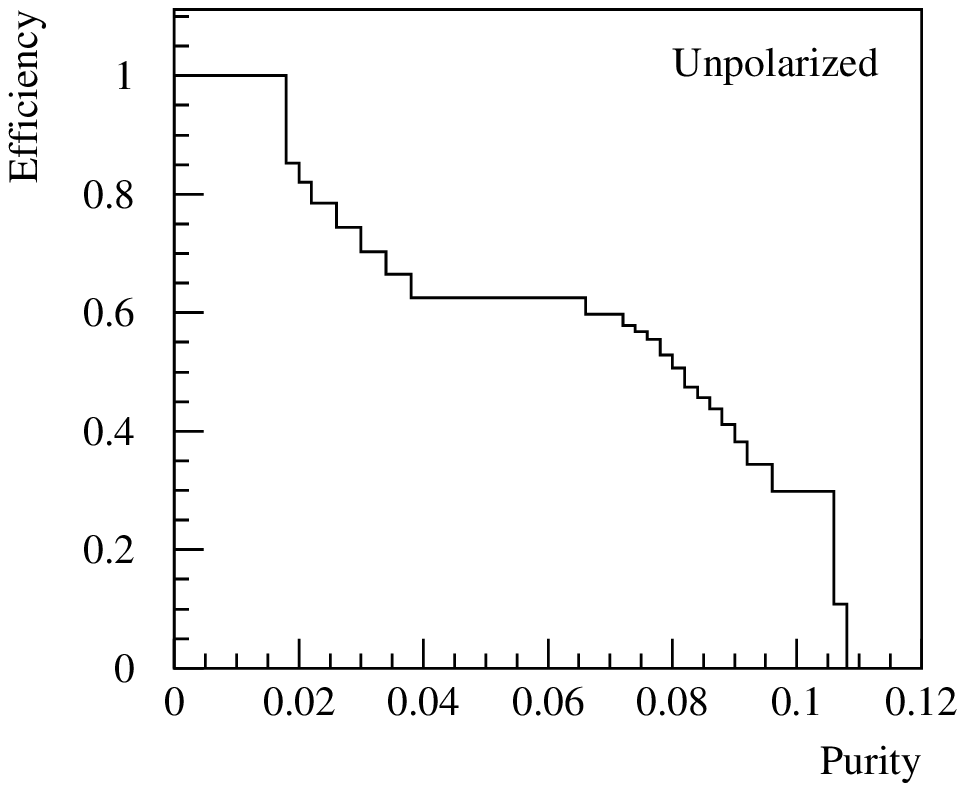,width=0.49\textwidth}}
\end{center}
\vspace*{-1.cm}
\caption{\label{fig:point5}
Left: Error bands and the corresponding error ellipse
as a function of \mt\ and \cost\ for $\sqrt s =500$~GeV
and ${\cal L}=500$~fb$^{-1}$.
For both plots $\mt=210$~GeV and $\cost=0.513$ are used.
Right: c-tagging efficiency vs. purity after event preselection
and SIMDET-4 detector simulation.}
\end{figure}

Figure~\ref{fig:point5} shows that the bounds on the scalar top 
mass at 210 GeV and $\cost=0.513$ increase.
We obtain $\Delta\mt=0.95$~GeV and $\Delta\cost=0.0125$. 
Based on the experience from direct searches at LEP, the systematic errors
on the event selection are less than 1\%; precise investigations require
the detailed detector layout and a full simulation.
The stop generator~\cite{asgenerator} has been interfaced 
with the SIMDET~\cite{simdet} 
simulation to allow an independent test of the detector simulation and related
systematic errors. Also with the SIMDET-4 simulation and based on an
implemented LCFI CCD detector simulation, a preliminary result of the 
c-tagging performance is presented.
Another uncertainty could arise from the luminosity measurement, 
the measurement of the polarization, 
and the theoretical uncertainty of the production cross section.
A high-luminosity linear collider with the capability of beam 
polarization has great potential for precision measurements 
in the scalar quark sector predicted by Supersymmetric theories.

\vspace*{-0.8cm}
\end{document}